\documentstyle[12pt]{article}
\begin{document}
\begin{center} 
{\bf A New Measure, Scale Invariance and See Saw Cosmology}
\end{center} 

\bigskip 
\begin{center}
E. I. Guendelman\footnote{Electronic address: guendel@bgumail.bgu.ac.il}    
\end{center}  
\bigskip 
\begin{center}
{\it Physics Department, Ben Gurion University, Beer Sheva 84105, Israel}
\end{center} 
\bigskip

\begin {abstract}
The cosmological constant problem and the possibility of obtaining
a see saw cosmological effect, where the effective vacuum energy
is highly suppressed by the existence of a large scale is investigated
in the context of scale-invariant, generally          
covariant theory. Scale invariance is considered  in          
the context of a gravitational theory where                                   
the action, in the first order formalism, is of the form $S =                 
\int L_{1} \Phi d^4x$ + $\int L_{2}\sqrt{-g}d^4x$ where $\Phi$ is a           
density built out of degrees of freedom independent of the metric.            
For global scale invariance, a "dilaton"                                      
$\phi$ has to be introduced, with non-trivial potentials $V(\phi)$ =          
$f_{1}e^{\alpha\phi}$ in $L_1$ and $U(\phi)$ = $f_{2}e^{2\alpha\phi}$ in      
$L_2$. In the effective Einstein frame, this leads to a non-trivial  
a potential (of the Morse type) for          
$\phi$ which has a flat region with energy density $f_{1}^{2}/4f_{2}$.
If the scale of $f_{1}$ is the electroweak scale, 
so that it is of the order of ${M_{EW}^{4}}$,  and the scale of $f_{2}$
is the Planck scale, so that it is of the order of ${M_{Pl}^{4}}$, 
then naturally we are led to a small vacuum energy for
the present universe of order
$M_{EW}^{8}/M_{Pl}^{4}$. This is exactly 
what is needed in the new approach to cosmic coincidences discussed by 
Arkani-Hamed et. al. . The model discussed does not have the problems usually
associated with scalar-tensor theories. 
\end {abstract}

\section{Scale Invariance with a New Measure}
The concept of scale invariance appears as an attractive possibility for a    
fundamental symmetry of nature. In its most naive realizations, such a        
symmetry is not a viable symmetry, however, since nature seems to have        
chosen some typical scales.                                                                       
Here we will find that spontaneously broken scale invariance 
can nevertheless be incorporated      
into realistic, generally covariant field theories. However, scale           
invariance has to be discussed in a more general framework than that of       
standard generally relativistic theories, where we must allow in the          
action, in addition to the                                                    
ordinary measure of integration $\sqrt{-g}d^{4}x$, another one,               
$\Phi d^{4}x$, where $\Phi$ is a density built out of degrees of freedom      
independent of the metric.                                                    
                                                                              
        For example, given 4-scalars $\varphi_{a}$ (a =                       
1,2,3,4), one can construct the density                                       
\begin{equation}                                                              
\Phi =  \varepsilon^{\mu\nu\alpha\beta}  \varepsilon_{abcd}                   
\partial_{\mu} \varphi_{a} \partial_{\nu} \varphi_{b} \partial_{\alpha}       
\varphi_{c} \partial_{\beta} \varphi_{d}                                      
\end{equation}                                                                
                                                                              
        One can allow both geometrical                                        
objects to enter the theory and consider$^1$                                  
\begin{equation}                                                              
S = \int L_{1} \Phi  d^{4} x  +  \int L_{2} \sqrt{-g}d^{4}x                   
\end{equation}                                                                

         Here $L_{1}$ and $L_{2}$ are                                         
$\varphi_{a}$  independent. There is a good reason not to consider            
mixing of  $\Phi$ and                                                         
$\sqrt{-g}$ , like                                                            
for example using                                                             
$\frac{\Phi^{2}}{\sqrt{-g}}$. This is because (2) is 
invariant (up to the integral of
a total
divergence) under the infinite dimensional symmetry                           
$\varphi_{a} \rightarrow \varphi_{a}  +  f_{a} (L_{1})$                       
where $f_{a} (L_{1})$ is an arbitrary function of $L_{1}$ if $L_{1}$ and      
$L_{2}$ are $\varphi_{a}$                                                     
independent. Such symmetry (up to the integral of a total divergence) is      
absent if mixed terms are present.                                            
                                                                              
        We will study now the dynamics of a scalar field $\phi$ interacting   
with gravity as given by the action (2) with$^{2,3}$                          
\begin{equation}                                                              
L_{1} = \frac{-1}{\kappa} R(\Gamma, g) + \frac{1}{2} g^{\mu\nu}               
\partial_{\mu} \phi \partial_{\nu} \phi - V(\phi),  L_{2} = U(\phi)           
\end{equation}                                                                
                                                                              
\begin{equation}                                                              
R(\Gamma,g) =  g^{\mu\nu}  R_{\mu\nu} (\Gamma) , R_{\mu\nu}                   
(\Gamma) = R^{\lambda}_{\mu\nu\lambda}, R^{\lambda}_{\mu\nu\sigma} (\Gamma)   
= \Gamma^{\lambda}_                                                           
{\mu\nu,\sigma} - \Gamma^{\lambda}_{\mu\sigma,\nu} +                          
\Gamma^{\lambda}_{\alpha\sigma}  \Gamma^{\alpha}_{\mu\nu} -                   
\Gamma^{\lambda}_{\alpha\nu} \Gamma^{\alpha}_{\mu\sigma}.                     
\end{equation}                                                                
                                                                              
        In the variational principle $\Gamma^{\lambda}_{\mu\nu}$,
$g_{\mu\nu}$, the measure fields scalars                                         
$\varphi_{a}$ and the  scalar field $\phi$ are all to be treated              
as independent variables.                                                     

        If we perform the global scale transformation ($\theta$ =             
constant)                                                                     
\begin{equation}                                                              
g_{\mu\nu}  \rightarrow   e^{\theta}  g_{\mu\nu}                              
\end{equation}                                                                
then (2), with the definitions (3), (4), is invariant provided  $V(\phi)$     
and $U(\phi)$ are of the                                                      
form                                                                          
                                                                                
\begin{equation}                                                              
V(\phi) = f_{1}  e^{\alpha\phi},  U(\phi) =  f_{2}                            
e^{2\alpha\phi}                                                               
\end{equation}                                                                
and $\varphi_{a}$ is transformed according to                                 
$\varphi_{a}   \rightarrow   \lambda_{a} \varphi_{a}$                         
(no sum on a) which means                                                     
$\Phi \rightarrow \biggl(\prod_{a} {\lambda}_{a}\biggr) \Phi \equiv \lambda   
\Phi $                                                                        
such that                                                                     
$\lambda = e^{\theta}$                                                        
and                                                                           
$\phi \rightarrow \phi - \frac{\theta}{\alpha}$. In this case we call the     
scalar field $\phi$ needed to implement scale invariance "dilaton".           
                                                                           
      Let us consider the equations which are obtained from                 
the variation of the $\varphi_{a}$                                            
fields. We obtain then  $A^{\mu}_{a} \partial_{\mu} L_{1} = 0$                
where  $A^{\mu}_{a} = \varepsilon^{\mu\nu\alpha\beta}                         
\varepsilon_{abcd} \partial_{\nu} \varphi_{b} \partial_{\alpha}               
\varphi_{c} \partial_{\beta} \varphi_{d}$. Since                              
det $(A^{\mu}_{a}) =\frac{4^{-4}}{4!} \Phi^{3} \neq 0$ if $\Phi\neq 0$.       
Therefore if $\Phi\neq 0$ we obtain that $\partial_{\mu} L_{1} = 0$,          
 or that                                                                      
$L_{1}  = M$,                                                                 
where M is constant. This constant M appears in a self-consistency            
condition of the equations of motion                                          
that allows us to solve for $ \chi \equiv \frac{\Phi}{\sqrt{-g}}$             

\begin{equation}                                                              
\chi = \frac{2U(\phi)}{M+V(\phi)}.                                            
\end{equation}

        To get the physical content of the theory, it is convenient to go     
to the Einstein conformal frame where                                         
\begin{equation}                                                              
\overline{g}_{\mu\nu} = \chi g_{\mu\nu}                                       
\end{equation}                                                                
and $\chi$  given by (7). In terms of $\overline{g}_{\mu\nu}$   the non       
Riemannian contribution (defined   as                                         
$\Sigma^{\lambda}_{\mu\nu} =                                                  
\Gamma^{\lambda}_{\mu\nu} -\{^{\lambda}_{\mu\nu}\}$                           
where $\{^{\lambda}_{\mu\nu}\}$   is the Christoffel symbol),                 
disappears from the equations, which can be written then in the Einstein      
form ($R_{\mu\nu} (\overline{g}_{\alpha\beta})$ =  usual Ricci tensor)        
\begin{equation}                                                              
R_{\mu\nu} (\overline{g}_{\alpha\beta}) - \frac{1}{2}                         
\overline{g}_{\mu\nu}                                                         
R(\overline{g}_{\alpha\beta}) = \frac{\kappa}{2} T^{eff}_{\mu\nu}             
(\phi)                                                                        
\end{equation}                                                                
where                                                                         
\begin{equation}                                                              
T^{eff}_{\mu\nu} (\phi) = \phi_{,\mu} \phi_{,\nu} - \frac{1}{2} \overline     
{g}_{\mu\nu} \phi_{,\alpha} \phi_{,\beta} \overline{g}^{\alpha\beta}          
+ \overline{g}_{\mu\nu} V_{eff} (\phi),                                       
V_{eff} (\phi) = \frac{1}{4U(\phi)}  (V+M)^{2}.                               
\end{equation}                                                                
        If $V(\phi) = f_{1} e^{\alpha\phi}$  and  $U(\phi) = f_{2} $  
as required by scale invariance, we obtain from (10)
\begin{equation}                                                              
        V_{eff}  = \frac{1}{4f_{2}}  (f_{1}  +  M e^{-\alpha\phi})^{2}        
\end{equation}

 Also a minimum is achieved at zero         
cosmological constant for the case $\frac{f_{1}}{M} < 0 $ at the point         
$\phi_{min}  =  \frac{-1}{\alpha} ln \mid\frac{f_1}{M}\mid $. Finally,        
the second derivative of the potential  $V_{eff}$  at the minimum is          
$V^{\prime\prime}_{eff} = \frac{\alpha^2}{2f_2} \mid{f_1}\mid^{2} > 0$        
if $f_{2} > 0$, so that a realistic scalar field potential, with           
massive excitations when considering the true vacuum state, is achieved in    
a way consistent with the idea of scale invariance.                           

\section{See-Saw Cosmology}            

        Since we can always perform the transformation $\phi \rightarrow      
- \phi$ we can                                                                
choose by convention $\alpha > O$. We then see from (11) that as
$\phi \rightarrow      
\infty, V_{eff} \rightarrow \frac{f_{1}^{2}}{4f_{2}} =$ const.                
providing an infinite flat region.                                                               

        If we  consider this model as a model for the present accelerated 
universe, we see that this flat region  can provide a long  
lived almost constant vacuum energy for a                                     
long period of time, which can be small if $f_{1}^{2}/4f_{2}$ is              
small. Such small energy                                                      
density will eventually disappear when the universe achieves its true         
vacuum state.                                                                 
                                                                              
        Notice that a small value of $\frac{f_{1}^{2}}{f_{2}}$   can be       
achieved if we let $f_{2} >> f_{1}$. In this case                             
$\frac{f_{1}^{2}}{f_{2}} << f_{1}$, i.e. a very small scale for the           
energy                                                                        
density of the universe is obtained by the existence of a very high scale     
(that of $f_{2}$) the same way as a small fermion mass is obtained in the     
see-saw mechanism$^{4}$ from the existence also of a large mass scale. In     
what follows, we will take $f_{2} >> f_{1}$.                                  

        If we assume that the only scales that can enter are the weak scale
$M_{EW}$ and associate this to $f_{1}$, 
so that it is of the order of ${M_{EW}^{4}}$,  and the Planck scale $M_{Pl}$ 
and associate this to $f_{2}$, 
so that it is of the order of ${M_{Pl}^{4}}$,  
we obtain that the energy density
 associated with the flat region is of the order 
$\frac{f_{1}^{2}}{f_{2}} = \frac{M_{EW}^{8}}{M_{Pl}^{4}}$. This is the 
fundamental requirement for the vacuum energy in a recent discussion 
of cosmic coincidences (see Refs.5  and contribution of H.Murayama to 
this conference).

\section{The Introduction of Fermions}

        So far we have studied a theory which contains the metric tensor      
$g_{\mu\nu}$, the measure fields $\varphi_{a}$ (a=1,2,3,4) and the            
"dilaton" $\phi$, which makes global scale invariance possible in a           
non-trivial way. All of the above fields have some kind of geometrical        
significance, but if we are to describe the real world, the list of fields    
and/or particles has to be enlarged.                                          
                                                                              
Taking, for example, the case of a fermion $\psi$, where the kinetic term     
of the fermion is chosen to be part of $L_1$                                  
\begin{equation}                                                              
S_{fk} = \int L_{fk} \Phi d^4 x                                               
\end{equation}                                                                

\begin{equation}                                                              
L_{fk} = \frac{i}{2} \overline{\psi} [\gamma^a V_a^\mu                        
(\overrightarrow{\partial}_\mu + \frac{1}{2} \omega_\mu^{cd} \sigma_{cd})     
- (\overleftarrow{\partial}_\mu + \frac{1}{2} \omega_\mu^{cd} \sigma_{cd})    
\gamma^a V^\mu_a] \psi                                                        
\end{equation}                                                                
there $V^\mu_a$ is the vierbein, $\sigma_{cd}$ =                              
$\frac{1}{2}[\gamma_c,\gamma_d]$, the spin connection $\omega^{cd}_\mu$ is    
determined by variation with respect to $\omega^{cd}_\mu$ and, for            
self-consistency, the curvature scalar is taken to be (if we want to deal     
with $\omega_\mu^{ab}$ instead of $\Gamma^\lambda_{\mu\nu}$ everywhere)       

\begin{equation}                                                              
R = V^{a\mu}V^{b\nu}R_{\mu\nu ab}(\omega),                                    
R_{\mu\nu ab}(\omega)=\partial_{\mu}\omega_{\nu ab}                           
-\partial_{\nu}\omega_{\mu ab}+(\omega_{\mu a}^{c}\omega_{\nu cb}             
-\omega_{\nu a}^{c}\omega_{\mu cb}).                                          
\end{equation}                                                                
                                                                              
Global scale invariance is obtained                                           
provided $\psi$ also transforms, as in                                        
$\psi \rightarrow \lambda ^{-\frac{1}{4}} \psi$. Mass term consistent with    
scale invariance exist,                                                       
\begin{equation}                                                              
S_{fm} = m_1 \int \overline{\psi} \psi e^{\alpha\phi/2} \Phi d^4x + m_2       
\int \overline{\psi} \psi e^{3\alpha\phi/2} \sqrt{-g} d^4 x.                  
\end{equation}

Let us discuss an interesting situation where                                 
coupling of the $ \phi $ field disappears and that the mass term              
becomes of a conventional form in the Einstein conformal frame.               
This is the case, when we study the theory for the limit                      
$\phi \rightarrow \infty$ . Then $U(\phi) \rightarrow \infty$ and             
$V(\phi) \rightarrow \infty$. In this case, taking                            
$m_1 e^{\alpha\phi/2} \overline{\psi}\psi$                                    
and $m_2 e^{3\alpha\phi/2} \overline{\psi}\psi$                               
much smaller than $V(\phi)$ or $U(\phi)$ respectively, therefore one can see  
that (7) is a good approximation and since                                    
also $M$ can be ignored in the self consistency condition (7)                 
in this limit, we get then,                                                   
$\chi = \frac{2f_2}{f_1} e^{\alpha\phi}$. If this used to express (15) in the
Einstein conformal frame (which must include also the transformation
$\psi ^\prime$ = $\chi ^{-\frac{1}{4}} \psi$ in order to get the canonical 
Einstein-Cartan form for the equations),         
we get $S_{fm} = 
m \int\sqrt{-\overline {g}}\overline{\psi} ^{\prime} \psi ^{\prime} d^4x$, 
where                                  

\begin{equation}                                                              
 m = m_1(\frac {f_1}{2f_2})^{\frac{1}{2}} +                                   
m_2(\frac {f_1}{2f_2})^{\frac{3}{2}}                                          
\end{equation}                                                          
 
Notice the interesting fact that the fermion mass is independent of the 
local value of the dilaton field $\phi$ in sharp contrast to Brans Dicke 
theories, where once we study the theory in the Einstein frame, particle 
masses are dependent on the local value of the dilaton field $\phi$. A 
difficult problem associated to Quintessential Models, i.e. the absence of
a 'fifth force' is naturally solved  by the same reason (all interaction
of dilaton are through the mass terms and they dissapear 
in the Einstein frame). 
One can show that particle masses are also constant (although a different
constant!) in the approximation of high fermion density$^{3}$.

\section{Discussion}

We have seen that a two measure theory leads naturally to a ground state 
with zero energy density. If scale invariance is implemented, a dilaton
potential appears which in addition to the minimum at zero, also has an 
infinite flat region. The energy density of this flat region naturally 
comes out to be just what is required for a succesful discussion of the cosmic 
coincidence problems. It is also remarkable that the resulting theory does
not have the problems of more standard scalar tensor theories (that is time 
dependent masses, fifth force) when the theory is anayzed in 
the infinite flat region of the dilaton potential.

\section*{Acknowledgments}
I would like to thank A.Kaganovich, H.Murayama and Y.Jack Ng 
for discussions.

\end{document}